\documentclass{aastex63}

\usepackage[utf8]{inputenc}
\usepackage{tablefootnote}
\usepackage{natbib}
\usepackage{amsmath}
\usepackage{threeparttable}
\usepackage{xspace}
\usepackage{graphicx}
\usepackage{multirow}
\usepackage{xcolor}

\definecolor{pink}{rgb}{1,0.1,.6}

\shorttitle{Flares, Rotation, and Planets of the AU Mic System}
\shortauthors{Gilbert et al.}

\begin{document}

\title{Flares, Rotation, and Planets of the AU Mic System from TESS Observations}

\correspondingauthor{Emily Gilbert}   
\email{emilygilbert@uchicago.edu}
\author[0000-0002-0388-8004]{Emily A. Gilbert}
\affiliation{Department of Astronomy and Astrophysics, University of
Chicago, 5640 S. Ellis Ave, Chicago, IL 60637, USA}
\affiliation{University of Maryland, Baltimore County, 1000 Hilltop Circle, Baltimore, MD 21250, USA}
\affiliation{The Adler Planetarium, 1300 South Lakeshore Drive, Chicago, IL 60605, USA}
\affiliation{NASA Goddard Space Flight Center, 8800 Greenbelt Road, Greenbelt, MD 20771, USA}
\affiliation{GSFC Sellers Exoplanet Environments Collaboration}

\author[0000-0001-7139-2724]{Thomas~Barclay}
\affiliation{University of Maryland, Baltimore County, 1000 Hilltop Circle, Baltimore, MD 21250, USA}
\affiliation{NASA Goddard Space Flight Center, 8800 Greenbelt Road, Greenbelt, MD 20771, USA}

\nocollaboration{2}{}

\author[0000-0003-1309-2904]{Elisa V. Quintana}
\affiliation{NASA Goddard Space Flight Center, 8800 Greenbelt Road, Greenbelt, MD 20771, USA}

\author[0000-0003-2918-8687]{Lucianne M. Walkowicz}
\affiliation{The Adler Planetarium, 1300 South Lakeshore Drive, Chicago, IL 60605, USA}

\author[0000-0002-5928-2685]{Laura D. Vega}
\affiliation{NASA Goddard Space Flight Center, 8800 Greenbelt Road, Greenbelt, MD 20771, USA}
\affiliation{Vanderbilt University, Department of Physics \& Astronomy, 6301 Stevenson Center Ln., Nashville, TN 37235, USA}

\author[0000-0001-5347-7062]{Joshua E. Schlieder}
\affiliation{NASA Goddard Space Flight Center, 8800 Greenbelt Road, Greenbelt, MD 20771, USA}

\author[0000-0003-3896-3059]{Teresa Monsue}
\affiliation{NASA Goddard Space Flight Center, 8800 Greenbelt Road, Greenbelt, MD 20771, USA}

\author[0000-0002-2078-6536]{Bryson Cale}
\affiliation{Department of Physics and Astronomy, George Mason University, Fairfax, VA, USA}

\author[0000-0003-2781-3207]{Kevin I. Collins}
\affiliation{Department of Physics and Astronomy, George Mason University, Fairfax, VA, USA}

\author[0000-0002-5258-6846]{Eric Gaidos}
\affiliation{Department of Earth Sciences, University of Hawai'i at Manoa, Honolulu, HI 96822 USA}

\author[0000-0001-8364-2903]{Mohammed El Mufti}
\affiliation{Department of Physics and Astronomy, George Mason University, Fairfax, VA, USA}

\author[0000-0003-4701-8497]{Michael Reefe}
\affiliation{Department of Physics and Astronomy, George Mason University, Fairfax, VA, USA}

\author[0000-0002-8864-1667]{Peter Plavchan}
\affiliation{Department of Physics and Astronomy, George Mason University, Fairfax, VA, USA}

\author{Angelle Tanner}
\affiliation{Mississippi State University, Department of Physics \& Astronomy, Hilbun Hall, Starkville, MS 39762, USA}

\author[0000-0001-9957-9304]{Robert A. Wittenmyer}
\affil{University of Southern Queensland, Centre for Astrophysics, West Street, Toowoomba, QLD 435 Australia}

\author[0000-0002-7424-9891]{Justin M. Wittrock}
\affiliation{Department of Physics and Astronomy, George Mason University, Fairfax, VA, USA}

\nocollaboration{14}{} 

\author{Jon M. Jenkins}
\affiliation{NASA Ames Research Center, Moffett Field, CA, 94035, USA}

\author[0000-0001-9911-7388]{David W. Latham}
\affiliation{Center for Astrophysics $\mid$ Harvard \& Smithsonian, 60 Garden St, Cambridge, MA, 02138, USA}

\author[0000-0003-2058-6662]{George R. Ricker}
\affiliation{Department of Physics and Kavli Institute for Astrophysics and Space Research, Massachusetts Institute of Technology, Cambridge, MA 02139, USA}

\author[0000-0003-4724-745X]{Mark E. Rose}
\affiliation{NASA Ames Research Center, Moffett Field, CA, 94035, USA}

\author[0000-0002-6892-6948]{S.~Seager}
\affiliation{Department of Physics and Kavli Institute for Astrophysics and Space Research, Massachusetts Institute of Technology, Cambridge, MA 02139, USA}
\affiliation{Department of Earth, Atmospheric and Planetary Sciences, Massachusetts Institute of Technology, Cambridge, MA 02139, USA}
\affiliation{Department of Aeronautics and Astronautics, MIT, 77 Massachusetts Avenue, Cambridge, MA 02139, USA}

\author[0000-0001-6763-6562]{Roland K. Vanderspek}
\affiliation{Department of Physics and Kavli Institute for Astrophysics and Space Research, Massachusetts Institute of Technology, Cambridge, MA 02139, USA}

\author{Joshua N. Winn}
\affiliation{Department of Astrophysical Sciences, Princeton University, 4 Ivy Lane, Princeton, NJ 08544,
USA}

\nocollaboration{7}{}

\begin{abstract}
AU Mic is a young ($\sim$24 Myr), pre-Main Sequence M~dwarf star that was observed in the first month of science observations of the Transiting Exoplanet Survey Satellite (TESS) and re-observed two years later. This target has photometric variability from a variety of sources that is readily apparent in the TESS light curves; spots induce modulation in the light curve, flares are present throughout (manifesting as sharp rises with slow exponential decay phases), and transits of AU Mic b may be seen by eye as dips in the light curve. We present a combined analysis of both TESS Sector 1 and Sector 27 AU Mic light curves including the new 20-second cadence data from TESS Year 3. We compare flare rates between both observations and analyze the spot evolution, showing that the activity levels increase slightly from Sector 1 to Sector 27. Furthermore, the 20-second data collection allows us to detect more flares, smaller flares, and better resolve flare morphology in white light as compared to the 2-minute data collection mode.
We also refine the parameters for AU Mic b by fitting three additional transits of AU Mic b from Sector 27 using a model that includes stellar activity. We show that the transits exhibit clear transit timing variations (TTVs) with an amplitude of $\sim$80 seconds. We also detect three transits of a 2.8 $R_\oplus$ planet, AU Mic c, which has a period of 18.86 days.

\end{abstract}



\section{Introduction} \label{sec:intro}

AU Mic (TIC 441420236, GJ 803) is a young \citep[24 $\pm$ 3 Myr,][]{bell15}, nearby \citep[9.9 $\pm$ 0.1 pc,][]{gaia2018}, bright \citep[$K$ = 4.5 mag,][]{stauffer2010} M1 pre-Main Sequence star. AU Mic is known for its frequent white light flares and its debris disk \citep{kalas04} that contains moving large-scale features \citep{Boccaletti15, boccaletti18}. The recent detection of a planet, AU Mic b, \citep{Plavchan2020} and a candidate planet c \citep{martioli2020} orbiting the star make this target even more intriguing. This pre-Main Sequence star provides us with valuable insight into how planetary systems are formed and develop and evolve in stellar infancy. For these reasons, AU Mic is an extremely well-studied star, and has been for decades. 

AU Mic has been theorized to be a young star \citep{Eggen1968,Kunkel1970,Wilson1970} and known to have flares since the late 1960s/early 1970s\footnote{AU Mic is listed by catalog number 395 in \citet{Eggen1968} who suggested that it was one of two that ``could represent stars that are still in the pre-main-sequence stage of their evolution".} \citep{Kunkel1970,Kunkel1973}. Its high amplitude photometric variability was first identified by \citet{Torres1972} who observed the star 50 times over a three month duration and suggested that dark surface spots were the cause of a 4.865 day modulation with a peak-to-peak amplitude of approximately 30\%. This photometric variability showed significant changes in amplitude during observations taken in the 1970s and 1980s \citep{Butler1987}, with the spot modulation similar to what we now see with two distinct spots or spot groups, apparently appearing in 1981 \citep{Rodono1986}.

The youth of AU Mic was verified via its membership in the $\beta$~Pictoris Moving Group \citep{zuckerman01, gagne18}. The star was first identified as having common galactic kinematics and a similar age to the eponymous $\beta$~Pictoris by \cite{Barrado99}. Subsequent study over the last two decades has firmly placed the star in the young stellar association and led to detailed estimates of its age using multiple methods. Consistent age estimates in the range of approximately 20 - 30 Myr have been found for the group (and thereby AU Mic) using isochronal analyses \citep{mamajek14, bell15}, lithium  depletion boundary studies \citep{binks14, malo14, macintosh15, shkolnik17}, as well as dynamical mass measurements and evolution model analyses of known binaries \citep{nielsen16, montet2015}. Here, we adopt the 24 $\pm$ 3 Myr isochronal age from \cite{bell15} for AU Mic. Recent summaries describing the history of the $\beta$ Pic group are available in \cite{brandt2014} and \cite{nielsen16}.

AU Mic continues to be studied and monitored by a wide variety of instruments to this day, including ongoing observation by the Transiting Exoplanet Survey Satellite (TESS). TESS is an all-sky mission designed to search for planets transiting nearby stars \citep{ricker14} using short-cadence, high-precision photometry. TESS started regular science operation on July 25, 2018 when it began observing in the Southern Ecliptic Hemisphere. TESS observed the Southern Ecliptic sky in 13 sectors, each $\sim$28 days long, totaling one year of observations. During this time, TESS observed a number of stars at 2-minute cadence and collected Full Frame Images (FFIs) every 30 minutes. After a year of observing in the south, TESS observed the Northern Ecliptic Hemisphere, in a similar pattern of 13 sectors at 2- and 30- minute cadences. TESS completed its primary mission in July 2020. Now in its extended mission, TESS once again observed in the Southern Ecliptic Hemisphere. TESS observed in a similar pattern to its first year of observations, observing the initial 13 sectors in order. In the extended mission, the data products have improved time sampling; each sector there are $>$500 20-second cadence targets, small cutouts called postage stamps are available at 2-minute cadence for 20,000 targets per sector, and FFI exposures are now captured every 10 minutes.

AU Mic b was discovered using observations from the first sector of TESS observations taken in July 2018 \citep{Plavchan2020}. The AU Mic light curve also showed significant variation as a result of stellar activity. Now, TESS has completed its third year of operations. AU Mic was reobserved in Sector 27 at 20-second cadence. This two year gap in observations plus the new higher cadence data make this system worthy of a reevaluation in order to refine the parameters for planet b and further characterize the activity of the host star. 

Herein, we present an overview of the data acquisition for both TESS observations (Section \ref{sec:data}). We then analyze the white light flares present in both the Sector 1 and Sector 27 TESS light curves and compare the results using both 2-minute and 20-second cadence data (Section \ref{sec:flares}). Next, we analyze the rotational spot modulation seen in both light curves (Section \ref{sec:spots}). We then present revised parameters for AU Mic~b by fitting all five transits seen by TESS (Section \ref{sec:transits}). Finally, we conclude by fitting three transits of AU Mic~c, a tentative detection of which was initially presented in \citet{Plavchan2020} (Section \ref{sec:planet-c}).

\section{Data and Observations} \label{sec:data}




TESS observed AU Mic twice: once in the first month of science operation (July 2018) and once in the third year of science (July 2020). The first observation of AU Mic\footnote{AU Mic was included on the 2-minute cadence list thanks to its inclusion on the following TESS Cycle 1 Guest Investigator (GI) programs: G011264/PI Davenport; G011175/PI Mann; G011180/PI Dressing; G011185/PI Davenport; G011266/PI Schlieder; G011176/PI Czekala; and G011239/PI Kowalski.} occurred during TESS Sector 1 observations using Camera 1, CCD 4. These observations were taken at 2-minute cadence and were completed during orbits 9 and 10 over the course of 27.9 days, with a 1.13 day interruption for data downlink between orbits, see Figure \ref{fig:sector1-fit}. Observations started on 2018 July 25 at 19:00:27 UTC and ended on 2018 August 22 at 16:14:51 UTC (TESS Julian Date (TJD)\footnote{TESS timestamps are Barycentric Julian Date - 2457000.} 1325.29278 to 1353.17778).

During the period of time from 2018 August 16 16:00 UTC to 2018 August 18 16:00 UTC (TJD 1347–1349), TESS had an improperly configured Fine Pointing mode calibration, contributing excess jitter and resulting in a significant noise increase. We filter out this data using data quality flags, resulting in some gaps in observation.

AU Mic was observed again\footnote{AU Mic was a target in 10 Cycle 3 GI proposals: G03263/PI Plavchan; G03273/PI Vega; G03141/PI Newton; G03063/PI Llama; G03205/PI Monsue; G03272/PI Burt; G03227/PI Davenport; G03228/PI Million; G03226/PI Silverstein; G03202/PI Paudel.} in TESS Sector 27 (TESS orbits 61 and 62), for 23.35 days of science operations (see Figures  \ref{fig:sector27-fit} and \ref{fig:sector27-20s-fit}). The observations began on 2020 July 05 at 18:31:16 UTC and ended on 2020 July 30 at 03:21:15 UTC (TJD 2036.27320 through 2060.64125). A 1.02 day gap for data downlink began on 2020 July 17 15:01:15 UTC. These data were collected at both 20-second and at 2-minute cadence. The 20-second cadence postage stamps consist of 10 co-adds of 1.98-second exposures.

We used the \texttt{lightkurve} \citep{lightkurve} package to download the AU Mic TESS light curves for Sectors 1 and 27 from the Mikulski Archive for Space Telescopes (MAST archive).  The Science Processing Operations Center (SPOC) processing the 2-min and 20-sec data to calibrate the image data, extract simple aperture photometry, and to identify and remove systematic effects from the resulting light curves \citep{jenkinsSPOC2016}. We chose to use the Presearch Data Conditioning (PDC\_SAP) light curves in our analysis \citep{Stumpe2012, Stumpe2014, smith2012}, which have been corrected for instrumental effects and for crowding. We used the `hardest' bitmask filter in \texttt{lightkurve}, and further filtered the data by removing NaNs. We confirmed by comparing these light curves to the unmasked light curves that no flares were removed in this process. This left us with light curves with 25.07, 23.29, and 22.57 days worth of observations for each Sector 1; Sector 27, 2-minute; and Sector 27, 20-second data respectively.

\section{Analysis of White Light Flares}\label{sec:flares}





We started by using an iterative Savitzky-Golay filter in order to flatten out the spot modulation in the raw light curves in preparation for flare detection. For the 2-minute cadence data, we iterated three times with a 10 hour window and fit a fourth order polynomial. With the 20-second cadence data, we iterated over five Savitzky-Golay fits, with a window length of five hours with a third order polynomial. Then we detected flares using a modified version of the \texttt{bayesflare} Python package \citep{pitkin14} that we have adapted to run on TESS short cadence data (both 2-minute and 20-second cadences). \texttt{bayesflare} uses Bayesian inference to detect flares in the data by running a sliding window over the data, comparing the data to the template, and ascribing an odds ratio to each data point. We used a flare template which models flares as a Gaussian rise ranging from 0 to 1.5 hours in duration followed by an exponential decay term ranging from half an hour to 3 hours in duration (modified slightly from \citet{pitkin14}) in order to optimize this software to detect small flares. This method of template matching allows us to differentiate between flares and instrumental noise and to accurately detect small flares. We then used \texttt{scipy}'s `find\_peaks' function in order to identify individual peaks within single complex flaring events using a 3$\sigma$ threshold. 

\begin{figure*}[h]
    \centering
    \includegraphics[width=.7\textwidth]{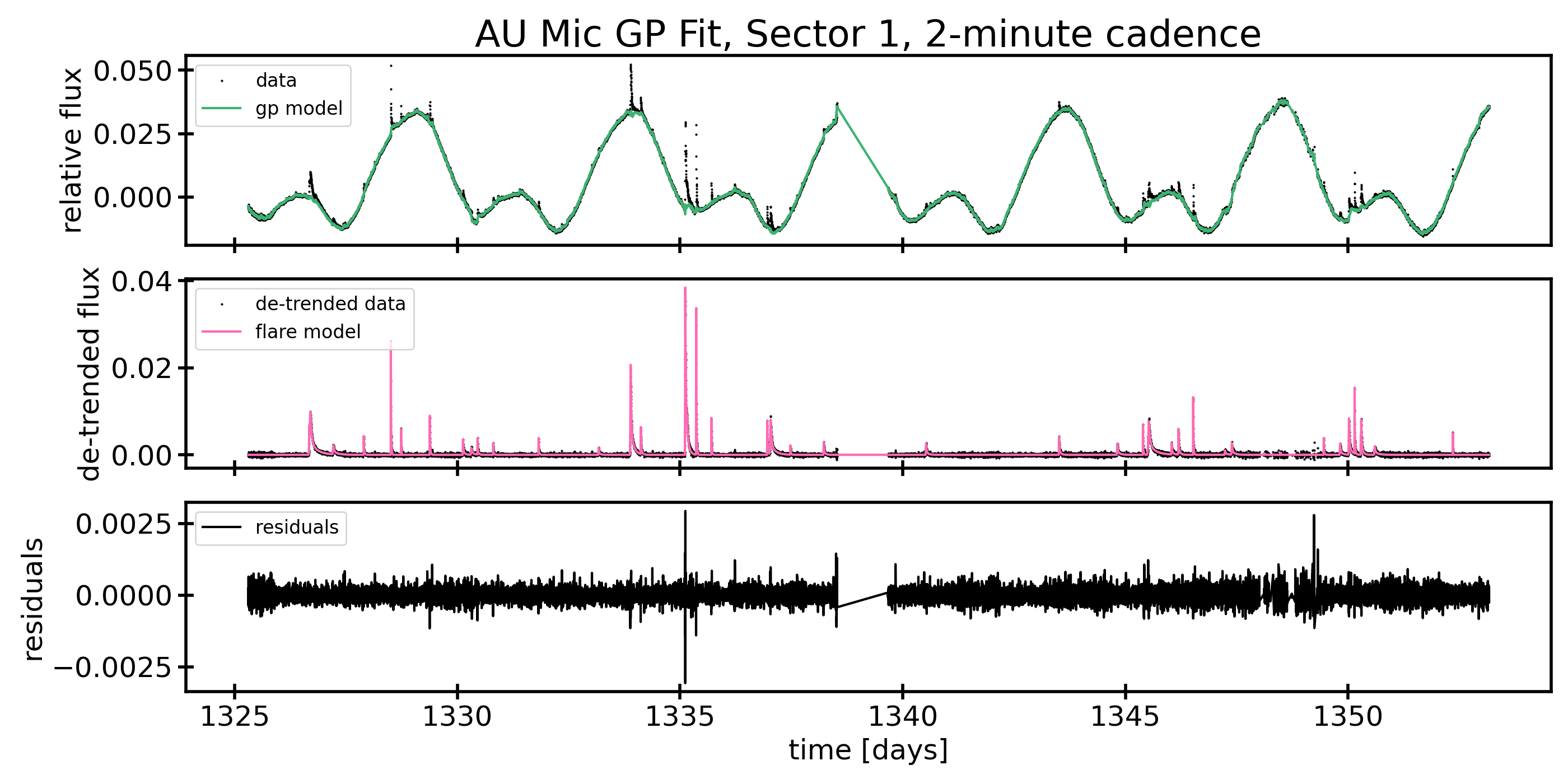}
    \caption{TESS Sector 1, 2-minute cadence observations of AU Mic. The top panel shows our GP spot model, the middle shows our flare model, and the bottom shows the residuals.}
    \label{fig:sector1-fit}
\end{figure*}

\begin{figure*}[h]
    \centering
    \includegraphics[width=.7\textwidth]{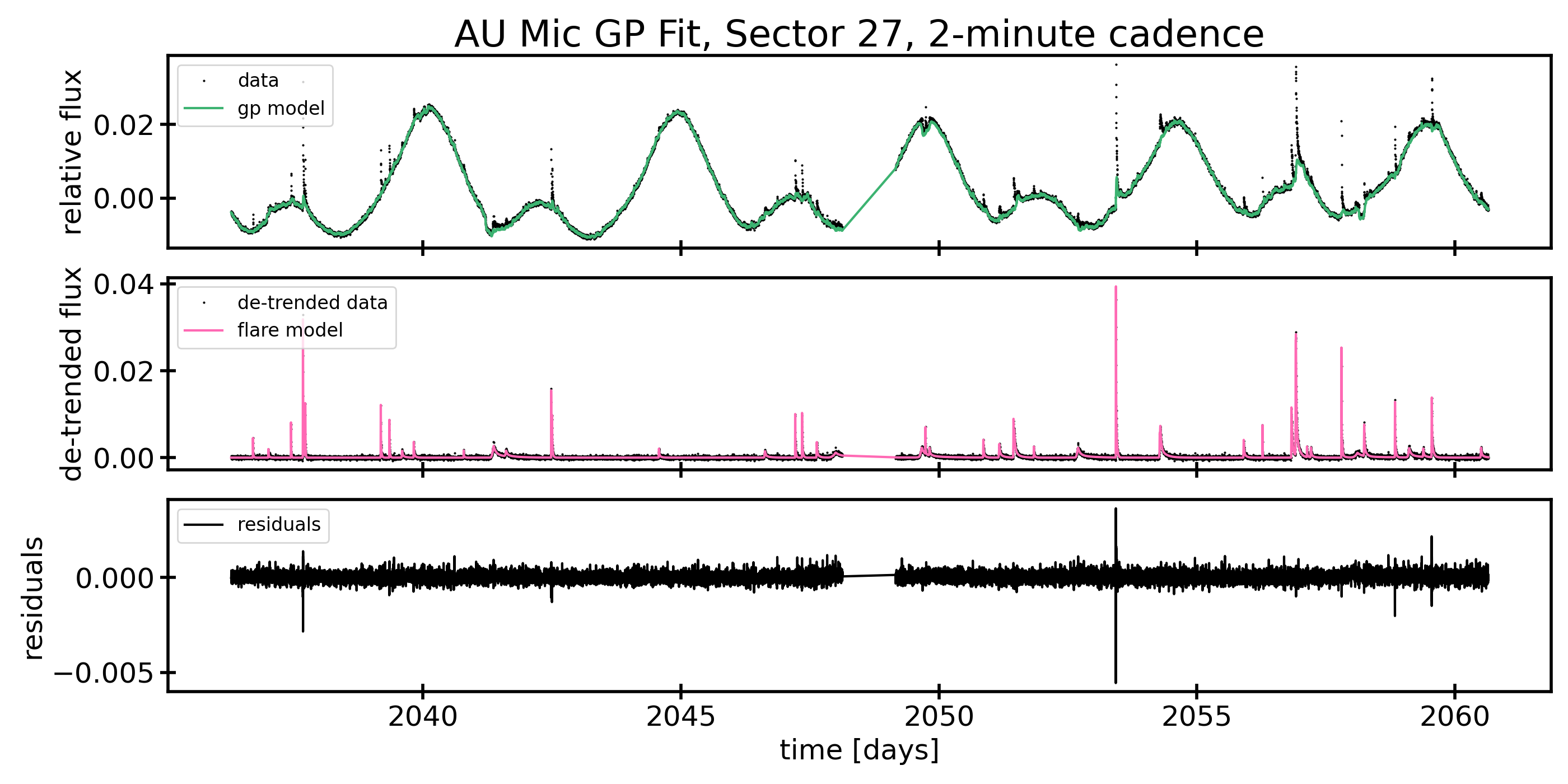}
        \caption{TESS Sector 27, 2-minute cadence observations of AU Mic. The top panel shows our GP spot model, the middle shows our flare model, and the bottom shows the residuals.}
    \label{fig:sector27-fit}
\end{figure*}

\begin{figure*}[h]
\centering
\includegraphics[width=.7\textwidth]{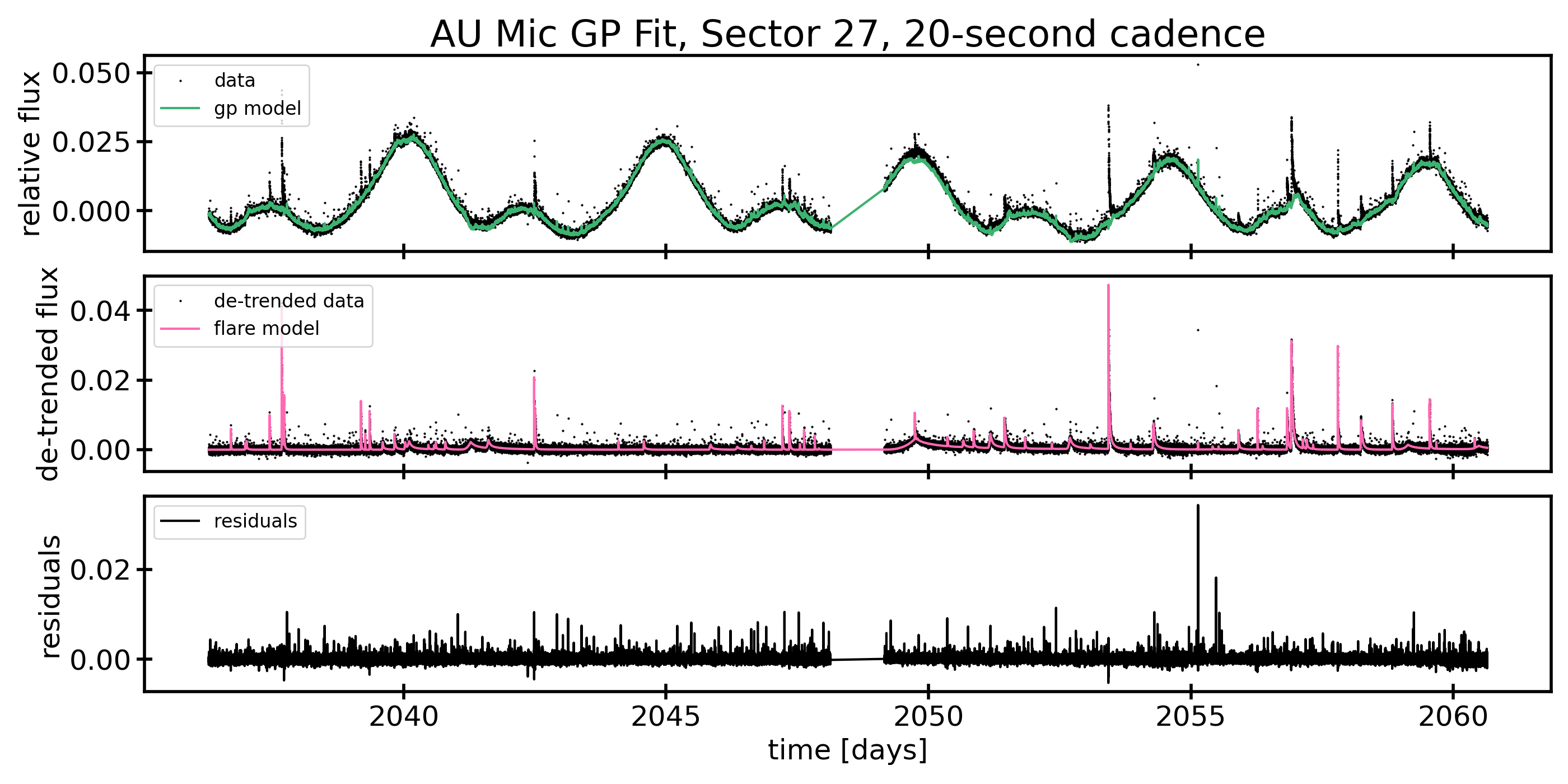}
\caption{TESS Sector 27, 20-second cadence observations of AU Mic. The top panel shows our GP spot model, the middle shows our flare model, and the bottom shows the residuals.}
    \label{fig:sector27-20s-fit}
\end{figure*}

Once flares are detected, \texttt{bayesflare} then returns a set of flare parameters: the peak time(s) of each flare, flare amplitudes, and the durations of each flare. We then used these parameters as initial guesses in a probabilistic model of the flares, with the flare shape based on the \citet{davenport14} flare profile. We returned to the original light curve in order to fit the spots and flares simultaneously. This combined fit requires a model implemented in Theano \citep{exoplanet:theano} to enable fast automatic differentiation of the function. Towards that end, we have written a small package called \texttt{xoflares} that can be use with \texttt{PyMC3}  \citep{exoplanet:pymc3} to model flares using the fast-rise, exponential decay profile of \citep{davenport14}. In order to fully model the light curve, we include a  Gaussian process (GP) component in the model simultaneously with the flare model \citep{exoplanet:foremanmackey17,exoplanet:foremanmackey18}. The GP kernel contains a jitter term describing excess white noise and a periodic term to capture spot modulation in the light curve. The best fitting models and residuals model can be seen in Figures \ref{fig:sector1-fit}, \ref{fig:sector27-fit}, and \ref{fig:sector27-20s-fit}.

After constructing our model, we sampled over the posterior. Due to the large number of parameters we are fitting, we elected to use \texttt{PyMC3}'s Automatic Differentiation Variational Inference algorithm \citep[ADVI,][]{advi2016} to approximate the posterior distribution of our model. We map the posterior distribution with 100,000 iterations. We then draw 3,000 samples from the posterior distribution to complete our fit.

\begin{figure*}[t]
\centering
\includegraphics[width=.75\textwidth]{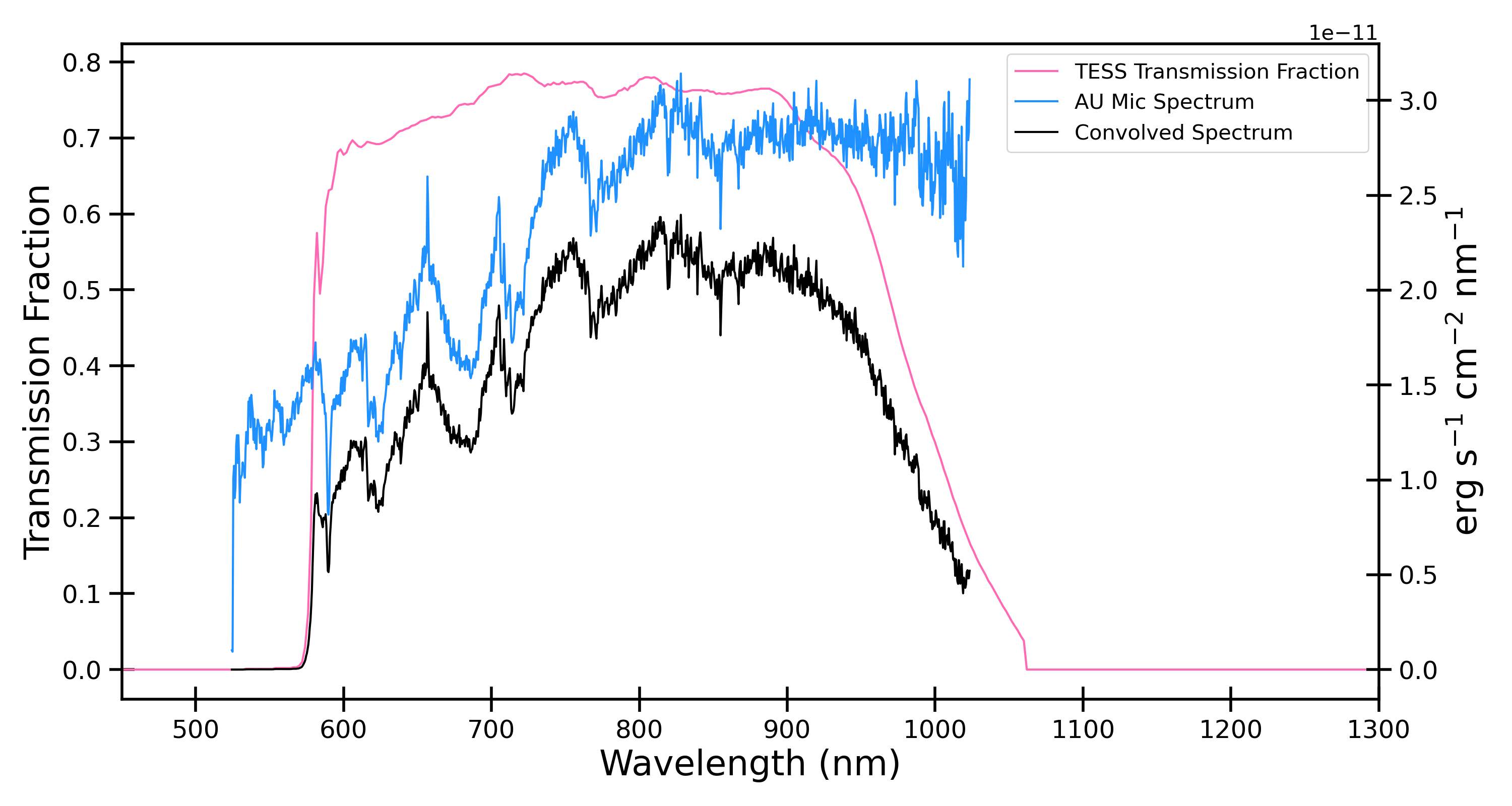}
\caption{We determine the reference energy for AU Mic by multiplying its spectrum by the TESS transmission function. The resulting convolved spectrum is shown in black. Integrating under the convolved spectrum gives us the quiescent energy emitted by AU Mic within the TESS bandpass.}
\label{fig:convolve}
\end{figure*}

From these samples, we were able to determine the distribution of model parameters (and therefore, energies) for each flare. To measure flare energies, we multiplied a flux-calibrated spectrum of AU Mic \citep[][private communication with Jamie Lomax, HST program GO-12512]{lomax2018} by the TESS response function \citep{Vanderspek2018b} and integrated the result across the TESS wavelength range, see Figure \ref{fig:convolve}.With this result, we found the energy per second per square centimeter emitted by a quiescent AU Mic detected by TESS to be: F$_{ref} = 1.0577*10^{-8} $ erg s$^{-1}$ cm$^{-2}$, which we subsequently used as a reference flux for the flares. Given the spectrum does not cover the full wavelength range of the TESS bandpass, energies are slightly underestimated.

\begin{figure*}[t]
\centering
\includegraphics[width=.95\textwidth]{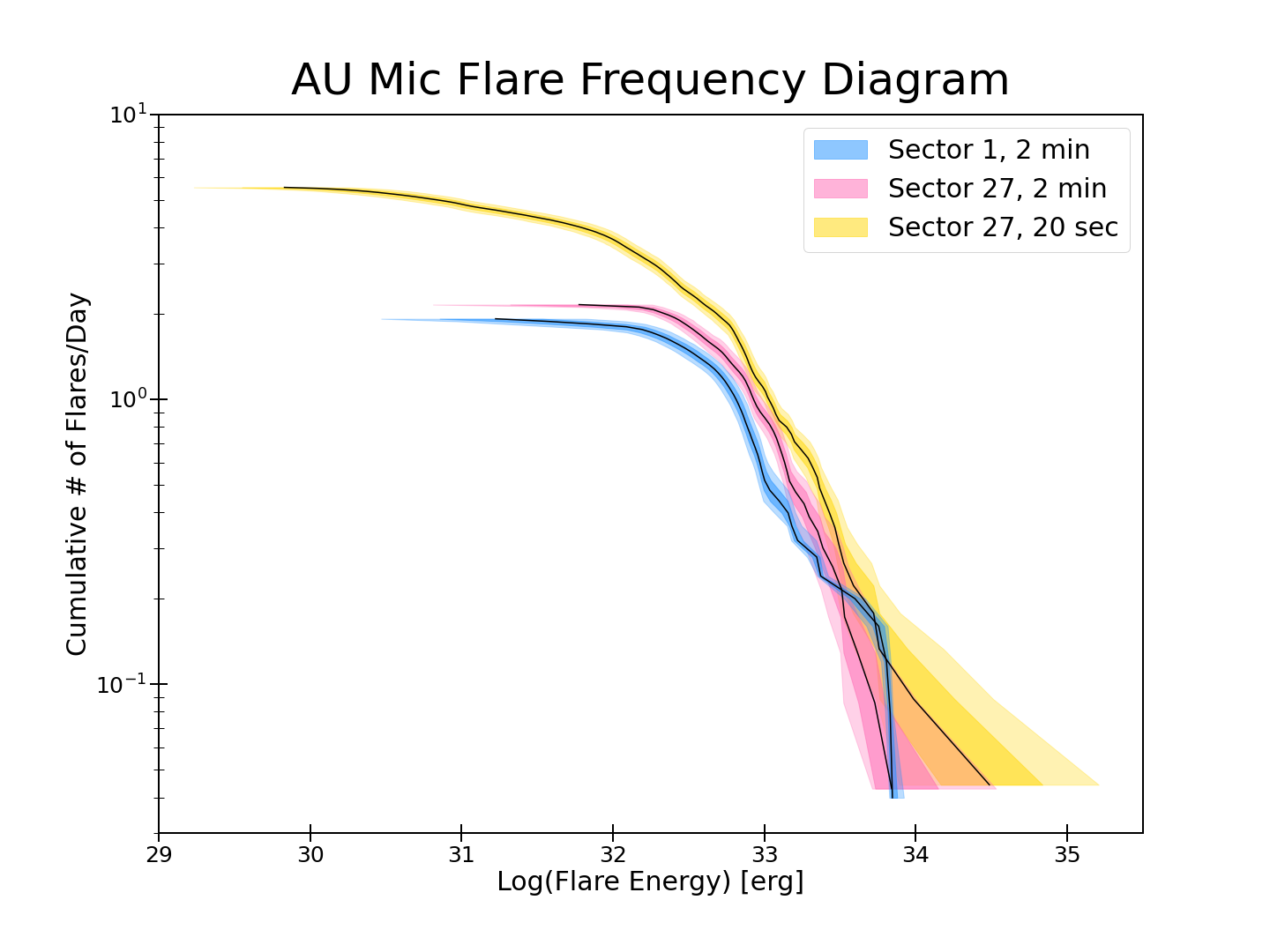}
\caption{The flare frequency distributions for AU Mic shows the flares modeled in both Sectors 1 and 27, as well as 2-minute and 20-second data for Sector 27 with the shaded regions showing the showing 1- and 2-$\sigma$ uncertainties. 20-second data significantly improves our ability to detect smaller (on order $ \sim 10^{32}$ erg) flares and resolve the full morphology of larger flares, resulting in larger flare energies. 2-minute data represents a only a lower limit on flare frequency and total energy.}
\label{fig:FFD-S1-S27}
\end{figure*}

Integrating under the curve of the flare model yields the equivalent duration of each flare. We were then able to determine the energy (within the TESS bandpass) emitted within each flare by scaling the equivalent duration by the reference energy and distance to AU Mic as follows:

\begin{equation}
    E_{abs} =  \int_{t_0}^{t_1} A(t) dt * F_{ref} * 4\pi d^2
\end{equation}

\noindent where the distance, obtained by GAIA DR2 parallax \citep{gaia2018}, is given by d = 9.7248 $\pm$ 0.004 pc, and A is the fractional amplitude of the flare relative to the star's quiescent flux as determined by our model. We integrated over all 3000 samples for each flare, which allowed us to determine uncertainties for each of the flare energies as shown in the flare frequency distribution presented in Figure \ref{fig:FFD-S1-S27}.

In order to objectively compare the activity levels from Sector 1 to Sector 27, we first investigated flare properties in 2-minute data for both sectors. We removed intervals of instrumental performance issues and calculated flare rates based on the length of time in the remaining data. In 25.07 days of observations from Sector 1, we detected 48 flares (1.855 flares/day), and in 23.29 days of observations in Sector 27, we detected 50 flares (2.147 flares/day). Sector 27 contains more low energy flares than Sector 1, see Figure \ref{fig:FFD-S1-S27}.

\begin{figure*}[h]
\centering
\includegraphics[width=.95\textwidth]{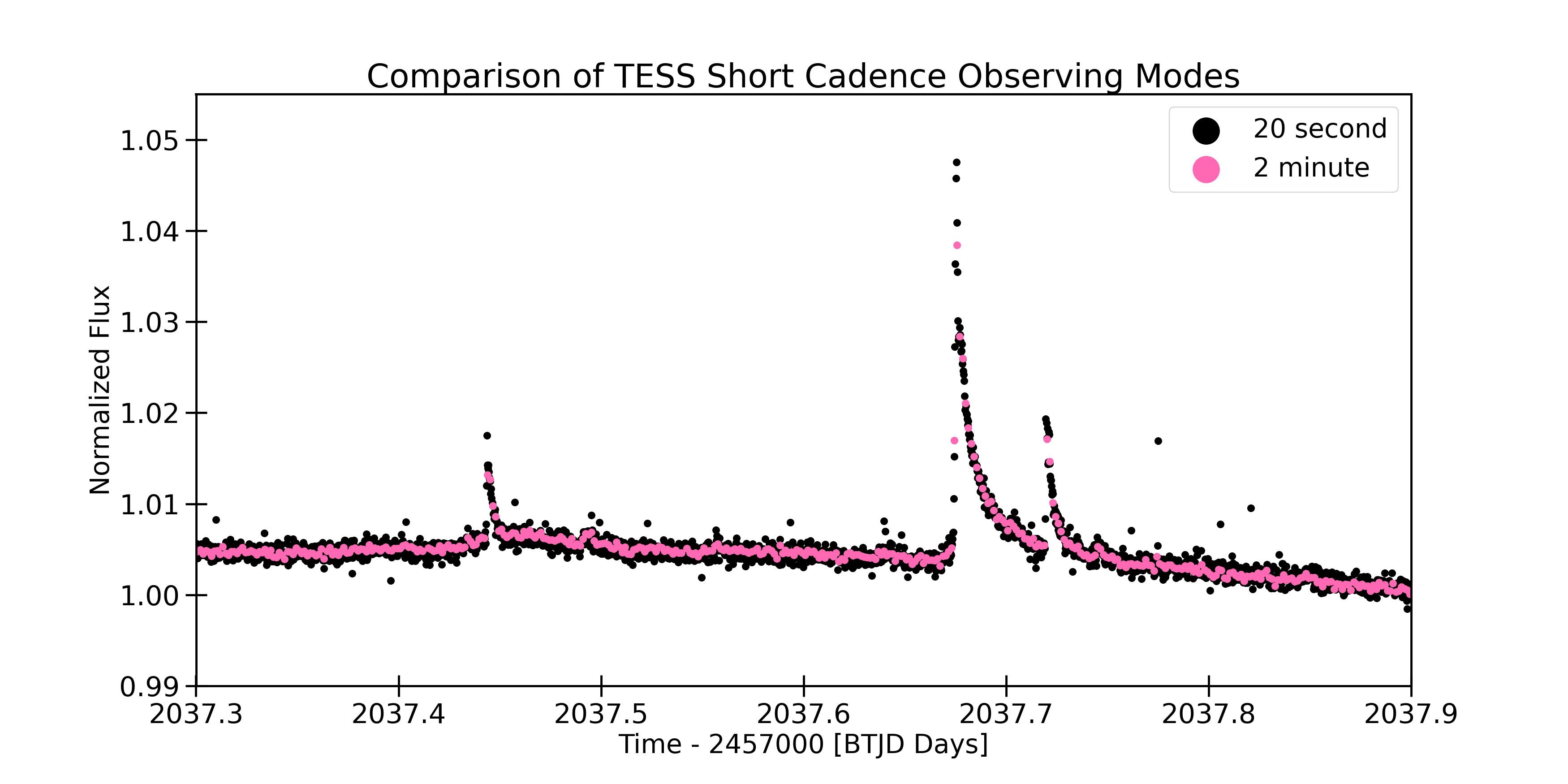}
\caption{20-second cadence data (black) allow us to better resolve flare peak amplitudes as compared to the 2-minute cadence data (pink).}
\label{fig:20s-comp}
\end{figure*}

We then looked at a comparison between the Sector 27 data obtained at both 2-minute and 20-second cadences. In the 22.57 days of 20-second data used, we detected 125 unique flares (5.54 flares/day). The 20-second data allows us to detect smaller flares than possible with the 2-minute data as shown clearly in Figure \ref{fig:FFD-S1-S27}. The 20-second data collection also enables us to fully resolve flare morphology, which may occur on timescales much faster than 2-minute data collection is capable of capturing, see Figure \ref{fig:20s-comp}. When we are unable to resolve the peak amplitude of the flare, the flare energies may be underestimated, as evidenced by the discrepancy in energies between the Sector 27 2-minute and 20-second data shown in Figure \ref{fig:FFD-S1-S27}.

\section{Spot Evolution}\label{sec:spots}

AU Mic has a distinct rotation pattern caused by the spots on the star seen in both sectors of TESS observations, see Figures \ref{fig:sector1-fit},  \ref{fig:sector27-fit}, and \ref{fig:sector27-20s-fit}. The spot pattern of AU Mic appears largely unchanged over the two years between TESS observations showing about 4\% peak-to-peak variability due to spots in both Sectors 1 and 27, although the amplitude decreased to approximately 3\% over the course of Sector 27. Given this persistent morphology, we expect that the same spots or spot groups are likely present during both TESS observations. The consistency in the spot pattern over two years shows that AU Mic has long-lived spot patterns. There does not appear to be any significant migration to differentially rotating latitudes on the star, although we are not especially sensitive to phase changes in the light curve as a result of our uncertainty in the rotation period, so cannot definitively rule changes out. Very long-lived spots have previously been seen on TESS targets, so this is consistent with observations of other M dwarf stars \citep[e.g.][]{Davenport2020,Robertson2020}.

\begin{figure}[h]
\centering
\includegraphics[width=.5\textwidth]{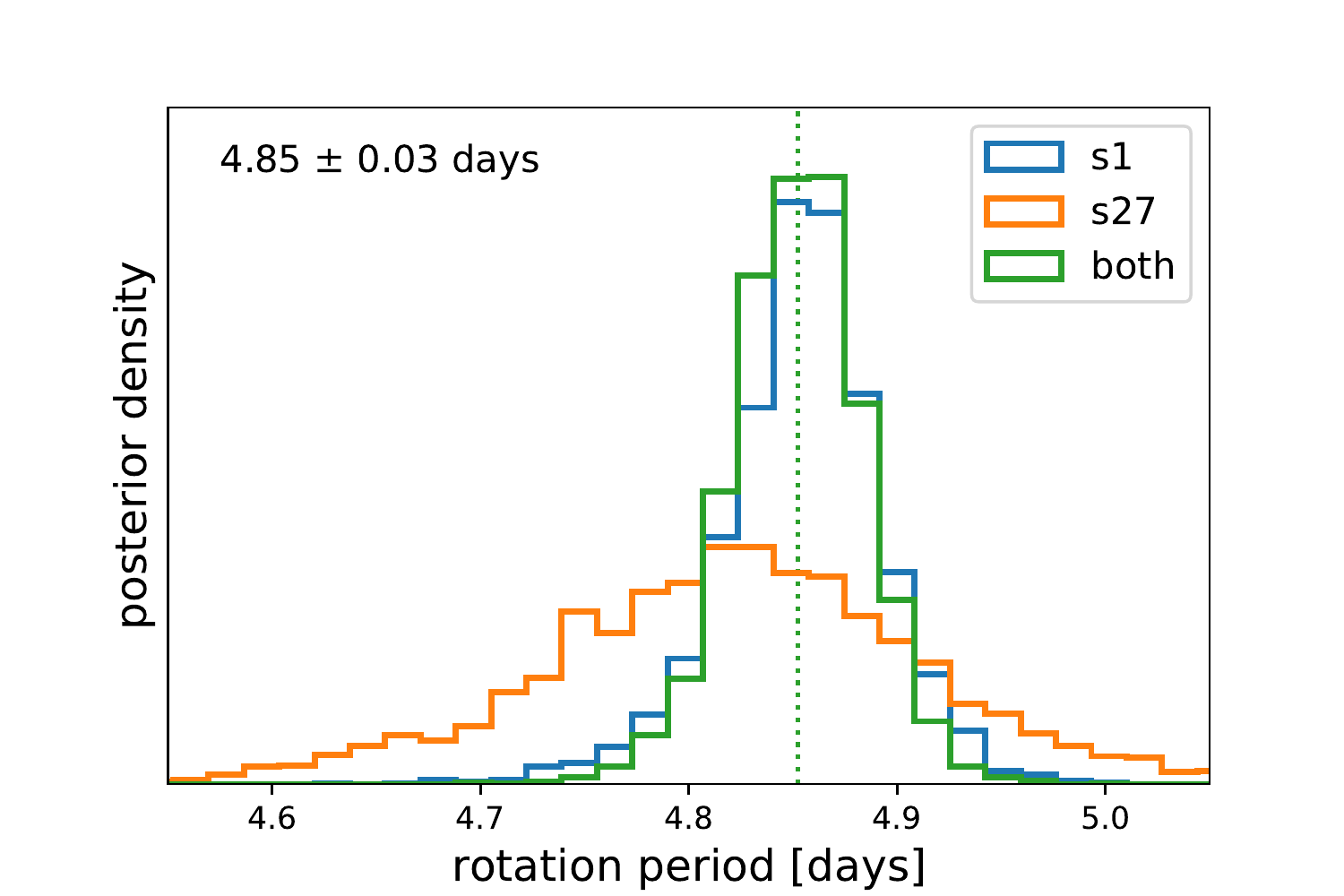}
\caption{Posterior model of the rotation of AU Mic, measured from the TESS photometric data. The Sector 1 data gives stronger constrains on the rotation period, primarily because there seems to be a little more light curve morphology change or phase drift in the Sector 27 data.}
\label{fig:rot_posterior}
\end{figure}

\begin{figure*}[t]
\centering
\includegraphics{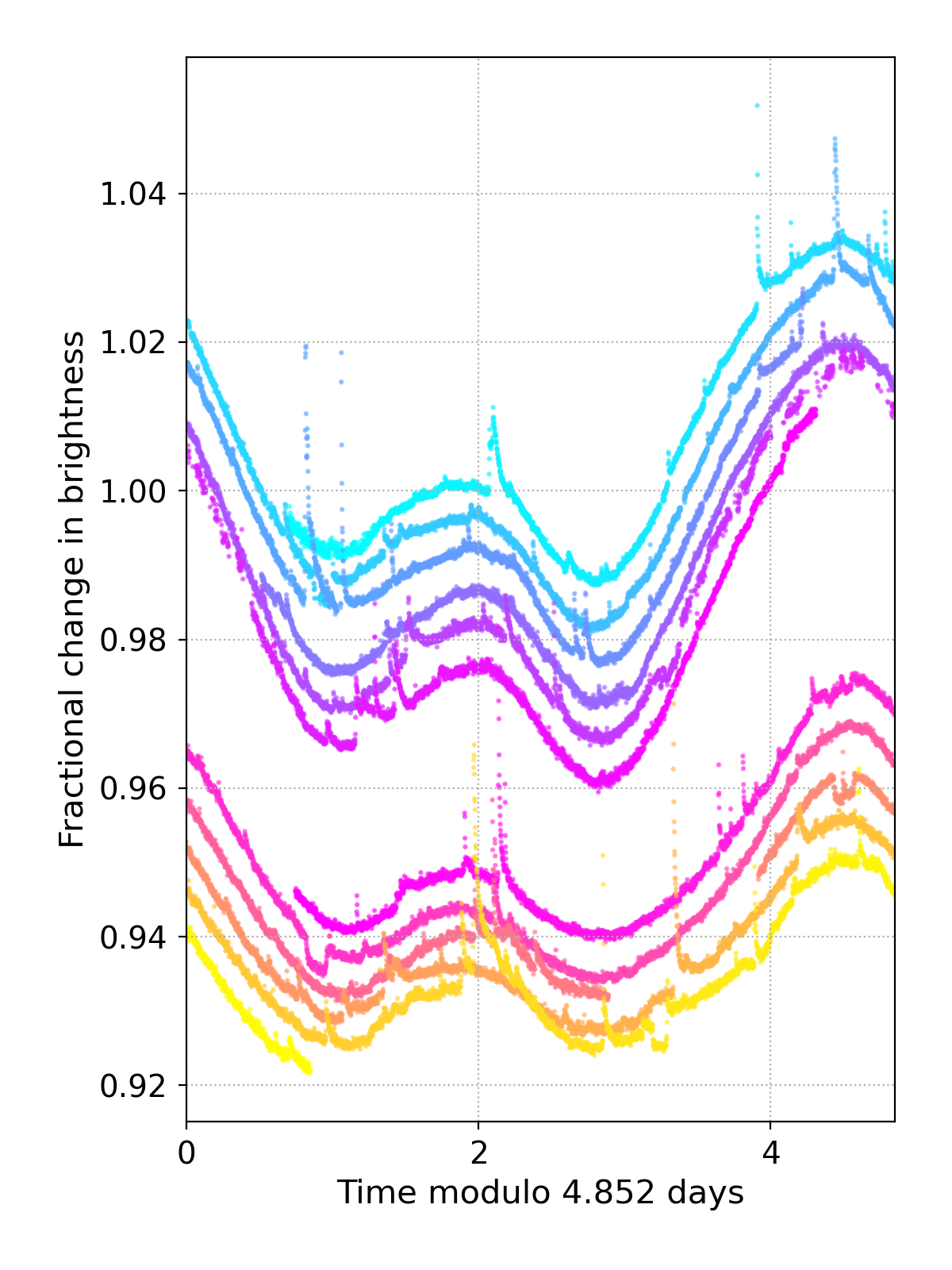}
\caption{Sector 1 and Sector 27 phase folded at the rotation period of the star, P = 4.85 days. Each phase offset vertically for clarity with time progressing downward. The overall shape of the light curve shows remarkable consistency over the two year observation baseline. The Sector 27 data has been shifted in phase to align with Sector 1 data.}
\label{fig:phasefold}
\end{figure*}

The significant spot modulation makes it straightforward for us to measure the rotation period of AU Mic. We elected to model this spots on their own using MCMC methods (separate from the variational inference modeling in Section \ref{sec:flares}) in order to obtain the most robust posteriors for the rotation period. Using the \texttt{exoplanet} package \citep{exoplanet:exoplanet} and \texttt{celerite} \citep{exoplanet:foremanmackey17,exoplanet:foremanmackey18}, we modeled the rotation period for each of the two TESS sectors using a periodic Gaussian Process kernel. As shown see Figure \ref{fig:rot_posterior} the period is well constrained. We measure the rotation period, derived from all datasets, to be P = 4.85 $\pm$ 0.03 days, we also get consistent periods when looking at each sector individually. The rotation period is less well constrained in Sector 27 than Sector 1, which we attribute to changes in the morphology of the light curve over rotation cycles in Sector 27 data. This may be due to spot migration and differential rotation on the surface of the star, or could be indicative of evolution in the spot pattern. In Figure \ref{fig:phasefold} we show the two light curves folded on the derived rotation period. While there are changes, the morphology over the two years of data changes very little.

As discussed in Section \ref{sec:intro}, historical observations show this same rotation period dating back 50 years. \citet{Torres1972} see a $\sim$30$\%$ variability due to spots in V band with a 4.865 day rotation period as observed from July to September, 1971. \citet{Rodono1986} show a series of light curves monitoring AU Mic in several different bands intermittently between 1971 and 1981. These light curves show variations in amplitude, morphology, and phase over time. The amplitudes have varied from $<$1\% to 30\% and therefore the 4\% we see today is not unusual. The double-peaked nature has not been observed continuously, but reports since the late 1980s all have the double-peaked structure.

\section{Refined Parameters for AU Mic b and Evidence for an Additional Massive Body in the System}\label{sec:transits}

Five transits of AU Mic b were observed by TESS. These data are particularly complicated to model because of the stellar variability from spots and flares which obscure transits and make fits less precise. Our goal with this section is to derive more precise parameters for the planet and to search for any transit timing variations (TTVs) in the TESS transits.

To model the transits, we build upon the work presented in \citet{Plavchan2020} that uses a transit model plus a Gaussian Process (GP) to model correlated noise in the light curve using the \texttt{exoplanet} package \citep{exoplanet:exoplanet} and dependencies, primarily \texttt{STARRY} \citep{exoplanet:luger18} and \texttt{celerite2} \citep{exoplanet:foremanmackey17,exoplanet:foremanmackey18}. 

The three additional TESS transits of AU Mic b, observed at 20-second cadence, are significant enhancements in the quantity of data we have available to characterize the planet. However, four out of the five total transits of AU Mic b observed by TESS have flares that occur during or immediately following the transit. Because the short term brightening during flares alters the shape of the transit, how flares are handled during transit fits affects the resulting planet parameters. The final transit observed has a particularly large flaring event that occurs during the final third of the transit and if not modeled correctly, gives the impression that the transit is significantly shorter than expected. In contrast to previous work by \citet{Plavchan2020}, who cut out the flares, we instead include them in our model with the goal of more precisely measuring the planet parameters and transit times. We identified times that showed a flare-like shape using the \texttt{bayesflare} model, and recorded the approximate peak time, amplitude, and duration. We then included in the light curve model a fast-rise exponential-decay component for each flare, following the flare profile described by \citet{davenport14}. Sampling of the model was performed using the No U-turn Sampler \citep[NUTS, ][]{NUTS} implemented in the \texttt{PyMC3} framework \citep{exoplanet:pymc3}. 

We used the \texttt{xoflares} Python package (described in Section \ref{sec:flares}) to model flares in and around the transits. In addition to including the flares in the model, we also allowed the time of the transits to vary so that we could get a constraint on potential TTVs. 

Since we are only interested in the transits, and including flares made our model computationally expensive, we opted to only include 12 hours of TESS data per transit, centered on the transit. Our model includes parameters for the stellar density, radius, two limb darkening parameters, and a mean value for each 12-hour window. We calculated the stellar density primarily based on \citet{Plavchan2020} values for stellar mass (0.50 $\pm 0.03$ M$_\odot$) and radius (0.75 $\pm 0.03$ R$_\odot$). However, to avoid over-constraining the models in the case of underestimated uncertainties we inflated the uncertainty on the density priors.
For stellar density, we used a natural log normal prior with a mean, $\mu$=0.5 log(g/cc), and standard deviation, $\sigma$=0.3; for radius we used a normal prior with mean, $\mu$=0.75 R$_\odot$, and standard deviation, $\sigma$=0.03, bounded at 0.1 and 2; and the limb darkening priors follow the \citet{exoplanet:kipping13} formalism for quadratic limb darkening.

We model the stellar variability as a GP representing a stochastically-driven, damped harmonic oscillator using \texttt{celerite2}, parameterized by the oscillation period and standard deviation. The oscillation period and standard deviation are consistent over the different transits. We also include a jitter parameter to describe excess white noise in the data (two jitter parameters in total, one for the 2-minute cadence Sector 1 data, and one for the 20-second cadence Sector 27 data).  The planet is parameterized by the planet-to-star radius ratio, impact parameter, orbital eccentricity, and periastron angle (parameterized as $e\sin{\omega}$ and $e\cos{\omega}$, and each transit by the transit time from which orbital period and transit epoch are derived). The priors on parameters are mostly weakly constraining lognormal, with orbital eccentricity following \citet{exoplanet:kipping13}, stellar radius being a normal with parameters from \citet{Plavchan2020}, transit times being zero-mean normals with a standard deviation of 30 minutes, and flare peak times being normals with with means from the best fit (t-2458692.97 = -362.52, -345.70, -345.58, 348.40, 356.76, 356.78, 365.27. 365.29) and a standard deviation of 0.003 days. The time offset in the light curve comes from setting time-zero in the model to be the central time in the whole time series. The planet-to-star radius ratio has a lognormal prior with a mean of 3 and standard deviation of 0.5. Impact parameter is uniformly sampled between 0 and 1 + the planet-to-star radius ratio. Periastron is uniform, but sampled to not see any discontinuities. Jitter and the two GP hyperparameters have inverse gamma priors.

\begin{figure*}[t]
\centering
\includegraphics[width=.75\textwidth]{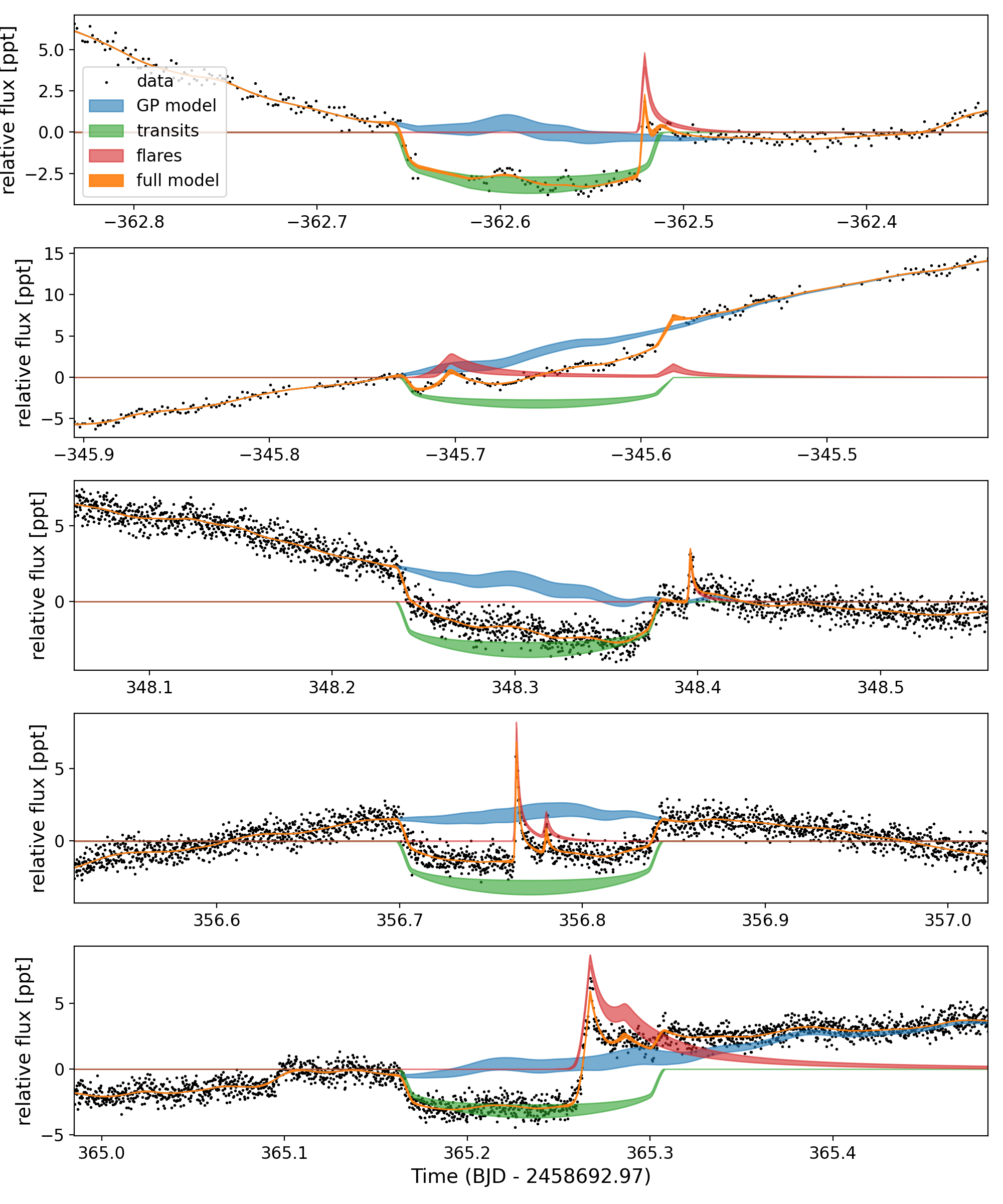}
\caption{Our model of the five TESS transits of AU Mic b and stellar flares that occur in and around the transit. The first two transits are Sector 1, 2-minute cadence, and the others are Sector 27, 20-second cadence.}
\label{fig:modeltransit}
\end{figure*}

\begin{figure*}[t]
\centering
\includegraphics[width=.75\textwidth]{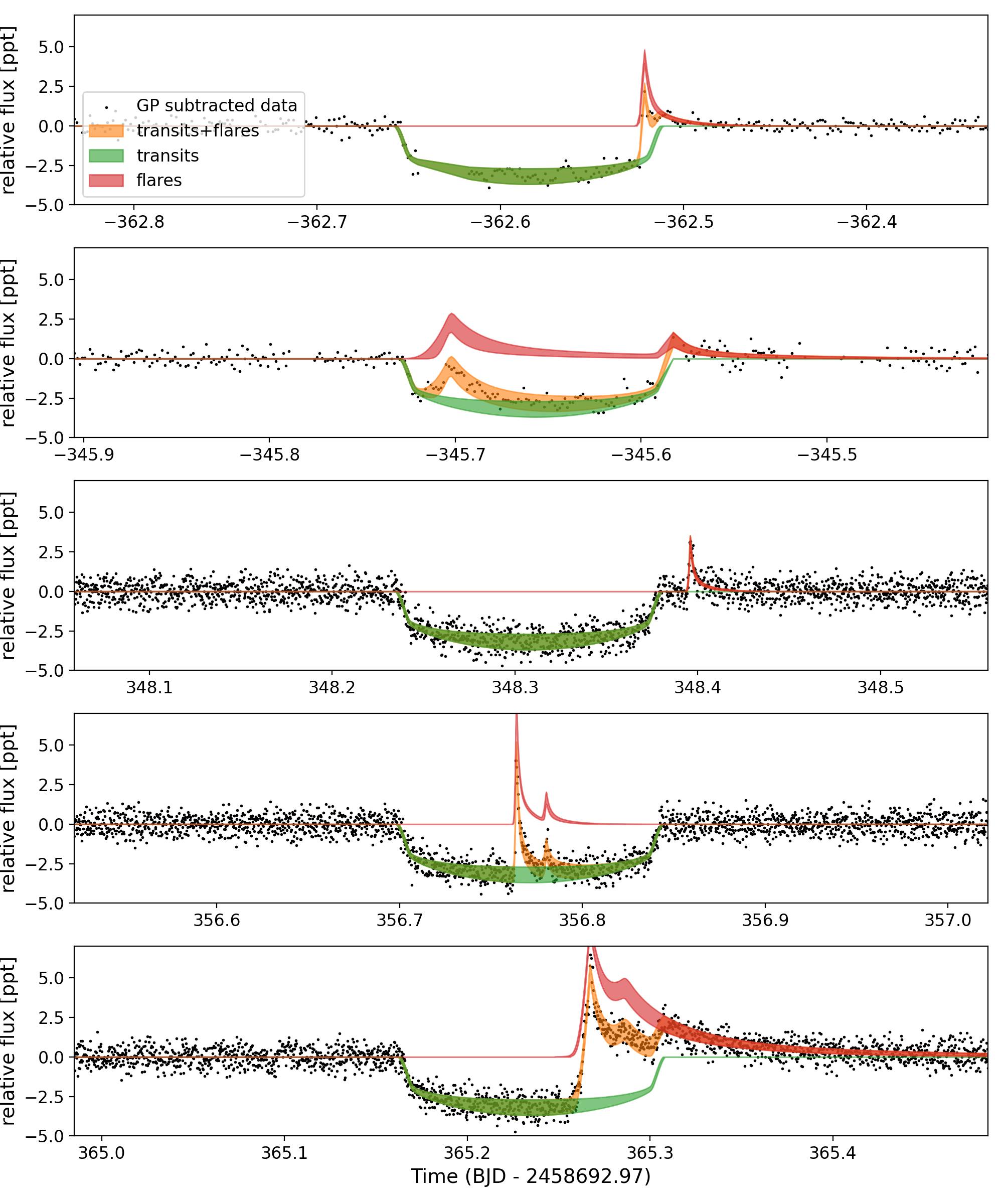}
\caption{The five TESS transits of AU Mic b with the the best fitting GP mean model subtracted from the data. The first two transits are Sector 1, 2-minute cadence, and the others are Sector 27, 20-second cadence.}
\label{fig:transits-gp}
\end{figure*}

We used the default sampler in \texttt{exoplanet}, with 4 chains, and an initial tuning phase of 5500 samples, followed by 2500 sample draw. The model converged, had Gelman–Rubin diagnostic values \citep{Gelman1992} of $<$1.001 with a minimum of 2000 independent samples in each model parameter. The model shown in data space is in Figure~\ref{fig:modeltransit}, where the blue curve shows the 80\% confidence range of the GP mean model, the green shows the transit and flare model, and the orange shows the sum of these two models. Figure~\ref{fig:transits-gp} has the GP mean model subtracted, showing the posterior constraints on the transit and flares. We find a best fitting orbital period of $8.463000\pm0.000006$ days and a planet radius of $4.2\pm0.2$ Earth-radii. We note that this period differs from the period reported by \citet{Plavchan2020} by about 18 seconds. While small, this is difference is highly significant. We attribute the difference to the earlier work not modeling the flares and measuring a transit duration that slightly too long.
The planet parameters are shown in Table~\ref{tab:aumicb}. We have reproduced the salient parameters derived in \citet{Plavchan2020} for comparison in Table~\ref{tab:aumicb-plavchan}. \citet{Plavchan2020} used one sector of TESS data coupled with one additional transit observed with \textit{Spitzer} as compared to two sectors worth of TESS observations used in this work. The stellar density and radius reported in Table~\ref{tab:aumicb} are primarily constrained by the prior rather than the data.

\begin{table}[]
    \centering
    \caption{AU Mic b model parameters}
    \begin{tabular}{cccc}
        Parameter&Value&+1-sigma & -1-sigma\\
        \hline
        Stellar density (g/cc) & 1.67 &  0.38 & 0.34 \\
        Stellar radius (R$_\odot$) & 0.750& 0.029 & 0.030 \\
        Orbital period (d) & 8.4630004 &0.0000058 & 0.0000060 \\
        T0 (BJD) & 2458330.39080 &0.00058 & 0.00057\\
        R$_\text{planet}$/R$_\text{star}$ & 0.0512 &0.0020 &0.0020 \\
        Impact parameter (b) & 0.26 &  0.13 & 0.17\\
        ecos$\omega$ & 0.01 & 0.38 & 0.36 \\
        esin$\omega$ & -0.05 & 0.22 & 0.20 \\
        Eccentricity & 0.12 & 0.16 & 0.08 \\
        Angle of periastron ($\omega$, radians) & -0.3 &  2.4 &   2.3 \\
        u$_1$ (limb darkening parameter) & 0.50 & 0.28 &  0.30 \\
        u$_2$ (limb darkening parameter) & 0.05 & 0.33 & 0.31\\
        Planet radius (R$_\oplus$) & 4.19 & 0.24 & 0.22 \\
        a/R$_\star$ & 18.5 & 1.3 & 1.4 \\
        a (AU) & 0.0644 & 0.0056 & 0.0054 \\
        Inclination (deg) & 89.18 & 0.53 & 0.45\\
        Transit duration (hours) & 3.56 & 0.60 & 0.46\\
    \end{tabular}
    
    \label{tab:aumicb}
\end{table}

\begin{table}[]
    \centering
    \caption{AU Mic b Parameters from \citet{Plavchan2020}}
    \begin{tabular}{cccc}
        Parameter&Value&+1-sigma & -1-sigma\\
        \hline
        Orbital period (d) & 8.46321 &0.00004 & 0.00004 \\
        T0 (BJD) & 2458330.39153 & 0.00070 & 0.00068 \\ 
        R$_\text{planet}$/R$_\text{star}$ & 0.00514  &0.0013 &0.0013 \\
        Impact parameter (b) & 0.16 &  0.14 & 0.11\\
        Eccentricity & 0.1 & 0.17 & 0.09 \\

        Planet radius (R$_\oplus$) & 4.203 & 0.202 & 0.202 \\
    \end{tabular}
    
    \label{tab:aumicb-plavchan}
\end{table}

\begin{figure}[t]
\centering
\includegraphics[width=.5\textwidth]{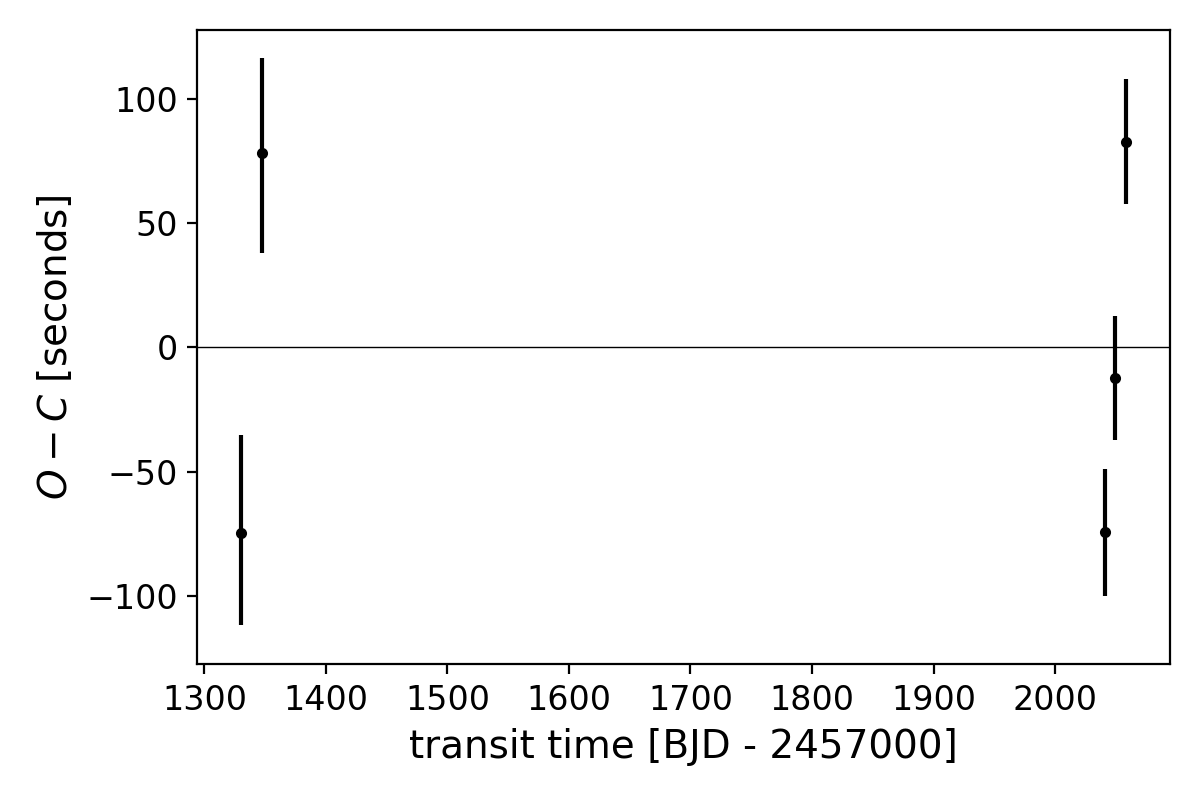}
\caption{Transit times as compared to the derived orbital period for all five transits of AU Mic b observed by TESS.}
\label{fig:ttvs}
\end{figure}

The transits are not exactly periodic; this is compelling evidence for TTVs in the TESS data. We find a TTV amplitude of approximately $\pm$80 seconds, shown in Figure~\ref{fig:ttvs}. With such a small TTV amplitude, we were only able to significantly detect the presence of TTVs with the aid of 20-second cadence and precise flare modeling. With only five data points and a two year gap between observations, it is difficult to draw too many conclusions from the presence of TTVs, aside from it providing evidence for at least one other massive body in the system, planet c.

\begin{table}[]
    \centering
    \caption{Times of AU Mic b transits observed by TESS.}
    \begin{tabular}{rr}
        Transit number&Transit times\\
        &(BJD -2457000) \\
        \hline
        0&$1330.38955\pm0.00065$\\
        2&$1347.31733\pm0.00066$\\
        84&$2041.28159\pm0.00036$\\
        85&$2049.74531\pm0.00036$\\
        86&$2058.20941\pm0.00037$\\
    \end{tabular}
    
    \label{tab:my_label}
\end{table}

\section{Detecting AU Mic c}\label{sec:planet-c}

\begin{figure*}
    \centering
    \includegraphics[width=.75\textwidth]{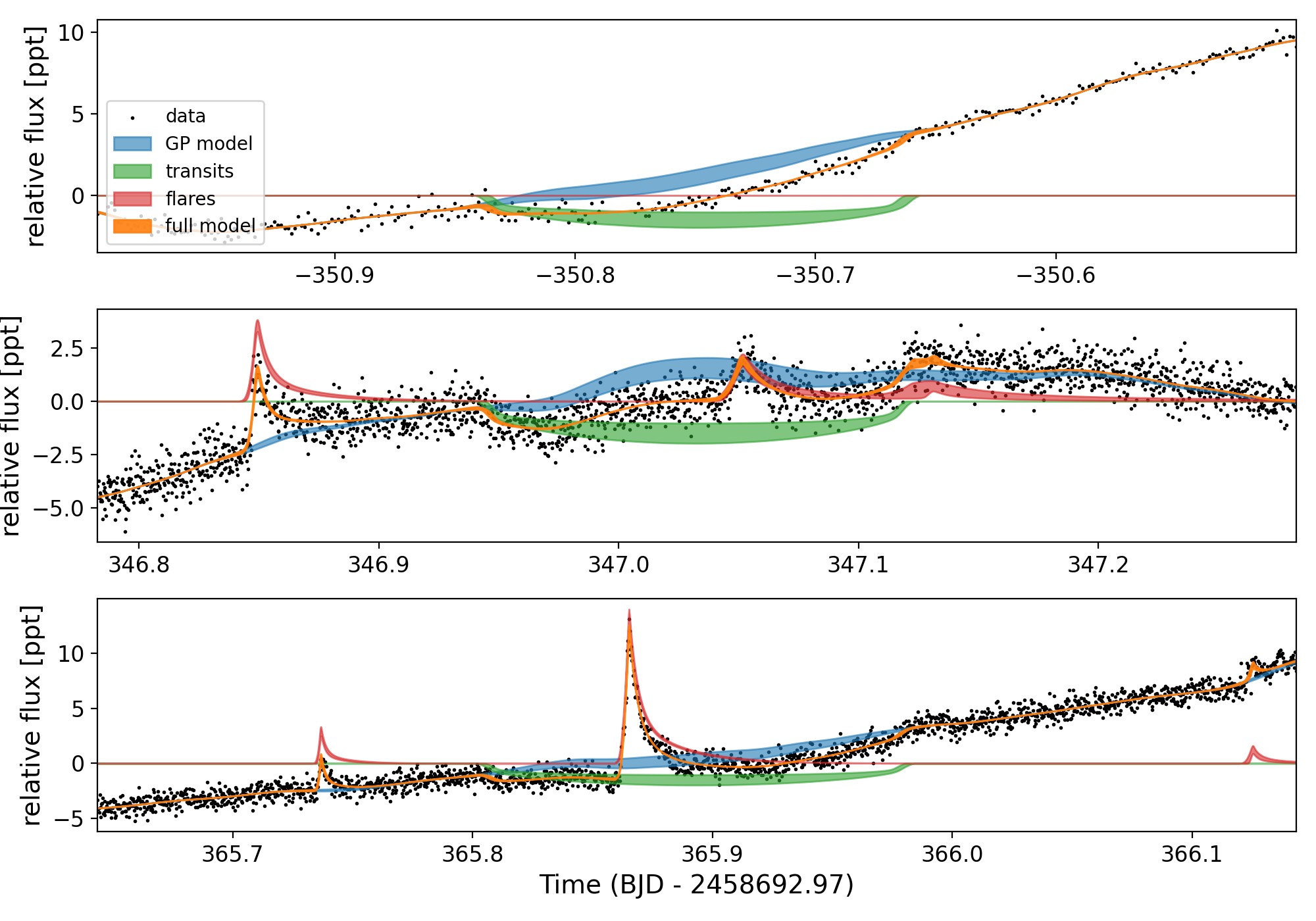}
    \caption{Three transits of AU Mic c. One from Sector 1 observed at 2-minute cadence, and two transits from Sector 27 observed at 20-second cadence.}
    \label{fig:planetc}
\end{figure*}

\begin{figure*}
    \centering
    \includegraphics[width=.75\textwidth]{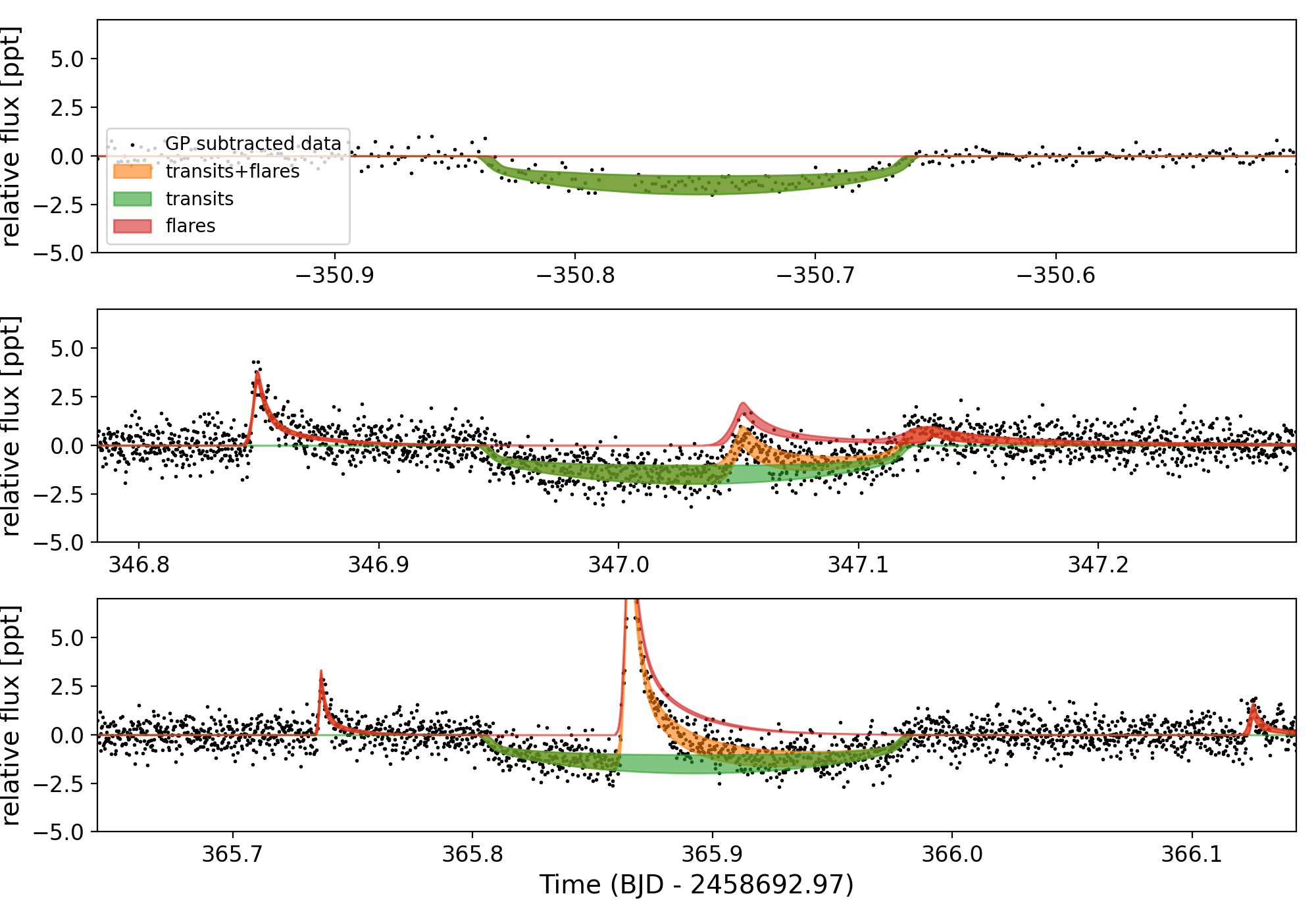}
    \caption{All three transits fits of AU Mic c, with the best fit GP mean model (stellar rotation and other long-term trends) subtracted off.}
    \label{fig:planetcflat}
\end{figure*}

\citet{Plavchan2020} reported the tentative detection of a second planet candidate in the AU Mic system with one potential transit even observed in Sector 1. With the additional data from Sector 27, we see evidence for two additional transits that are not associated with planet b. The duration and depth of the previous event and the two new events are consistent, and the spacing between them implies an orbital period of approximately 18.9 days. We modeled the transits of candidate planet c using the same method described in Section~\ref{sec:transits}, accounting for the transit timing variations shown in Figure~\ref{fig:ttvs}. The model parameters are provided in Table~\ref{tab:planetc} and the transit models are shown in Figures~\ref{fig:planetc} and \ref{fig:planetcflat}.

AU Mic c has a radius of $2.8\pm0.30$ R$_\oplus$. Flares occurred during two of the three transits, with a 1.4\% amplitude flare occurring close to the center of the final transit. By modeling these flares simultaneously with the data, we can derive an accurate radius for the planet. We do note that the radius is fairly uncertain, and this is because the model true transit depth is complicated by the uncertainty in flare amplitudes and durations. For both planets, we detect small, but significant eccentricities. 
AU Mic c has an orbital period of 2.23x that of AU Mic b, not especially close to any mean motion resonances. Nevertheless, it is possible that AU Mic c is causing the TTVs seen in the AU Mic b transit times (see Section~\ref{sec:martioli-comparison}). Further observations with \emph{Spitzer} provide independent confirmation of TTVs, which will be the subject of a subsequent paper (Wittrock et al., in prep).

\begin{table}[]
    \centering
    \caption{AU Mic c model parameters}
    \begin{tabular}{cccc}
        Parameter&Value&+1-sigma & -1-sigma\\
        \hline
        Orbital period (d) & 18.858982 &0.000053 & 0.000050 \\
        T0 (BJD) &2458342.2239 0.0017 0.0019\\
        R$_\text{planet}$/R$_\text{star}$ & 0.0340 &0.0034 &0.0033 \\
        Impact parameter (b) & 0.30 &  0.21 & 0.20\\
        ecos$\omega$ & -0.0 & 0.37 & 0.37 \\
        esin$\omega$ & -0.05 & 0.24 & 0.23 \\
        Eccentricity & 0.13 & 0.16 & 0.09 \\
        Angle of periastron ($\omega$, radians) & -0.3 &  2.5 &   2.2 \\
        Planet radius (R$_\oplus$) & 2.79 & 0.31 & 0.30 \\
        a/R$_\star$ & 31.7 2.6& 2.7 \\
        a (AU) & 0.110 & 0.010 & 0.010 \\
        Inclination (deg) & 89.39 &0.40 &0.38\\
        Transit duration (hours) & 4.42 &0.92& 0.67\\
\\
    \end{tabular}
    
    \label{tab:planetc}
\end{table}

\begin{table}[]
    \centering
    \caption{Times of AU Mic c transits observed by TESS.}
    \begin{tabular}{rr}
        Transit number&Transit times\\
        &(BJD -2457000) \\
        \hline
        0&$1342.2240\pm0.0018$\\
        37&$2040.0054\pm0.0011$\\
        38&$2058.8663\pm0.0011$\\
    \end{tabular}
    
    \label{tab:my_label}
\end{table}

\section{Discussion}\label{sec:martioli-comparison}

AU Mic presents similar, but not identical, levels of activity across the two sectors of TESS observations. The two year gap in observations between Sectors 1 and 27 allows us a brief glimpse into the evolution of activity on AU Mic over time. The flare rate increases slightly, but not significantly, from Sector 1 (1.855 flares/day) to Sector 27 (2.147 flares/day) as observed in 2-minute cadence data in each. However, the spot pattern is relatively consistent, with decreasing amplitude in Sector 27 (dropping from 4.3\% variability to 3.4\%). From our (unresolved) perspective, it is unclear if spots are changing, disappearing, or perhaps appearing more uniformly over the surface of the star, so we are unable to definitively say what may be causing the decrease in amplitude in the variability. Flares however, are a better indicator of the level of activity on a star \citep{notsu2013, doyle2020, feinstein2020}. On the sun, flare rates increase with spot complexity, so the higher flare energies are perhaps indicative of activity cycles occurring on AU Mic similar to those we see on the Sun where monthly flare rates can vary by a factor of 20 or more \citep{Aschwanden2012}. 

AU Mic is an extremely well-studied system that many groups have investigated. In particular, \citet{martioli2020} also looked at the TESS data included in this analysis. \citet{martioli2020} find planet b to be 4.07 $\pm 0.17 $R$_\oplus$, consistent with our results. They conduct a similar fit of the potential planet c transits, determining the planet candidate to be 3.24 $\pm 0.16 $R$_\oplus$ with a period of 18.86 days. While our results agree on the planet period, we find the radius of planet c to be significantly smaller at 2.79 $\pm0.3$ R$_\oplus$ as a result of differing modeling methods. In their analysis, \citet{martioli2020} find no evidence of transit timing variations, whereas we see evidence for TTVs on the order of $\sim$80 seconds. 

We used \texttt{TTV2fast2furious} \citep{Hadden2018}, the derived planet parameters, and planet masses estimated using \texttt{Forecaster} \citep{forecaster} to infer expected TTV amplitudes driven by interactions between planets b and c. The estimated amplitudes from the software are of order four minutes for each planet, which is at the same level of TTV amplitude we measure in the TESS data. While the number of transits we have is fairly limited and therefore a fully analysis of TTVs is not appropriate for the data presented here, the observed TTVs would seem to be consistent with interactions between AU Mic b and c.

\section{Conclusions}

AU Mic was observed twice by TESS: once in Sector 1 at 2-minute cadence and once in Sector 27 at 20-second cadence as well as 2-minute cadence. The introduction of the new 20-second data collection mode coupled with an additional sector of observations with TESS has enabled us to improve transit fits for AU Mic b and study the activity of AU Mic in unprecedented detail.  We detected and modeled flares in all available sectors of data, and show that AU Mic has grown slightly more active over the two year baseline from Sector 1 to Sector 27. Furthermore, the 20-second data collection mode allows us to detect more flares at smaller energies as well as resolve white light flare morphologies better than the 2-minute cadence alone. While the white light flare rate increases slightly from Sector 1 to Sector 27, the spot pattern of the star remains largely unchanged, indicating the likely persistence of a particular set of spots at preferred latitudes. 

We also used 3 additional transits of planet b observed by TESS in Sector 27 at 20-second cadence in order to refine the original planet fits as described in \citet{Plavchan2020} and report the detection of TTVs. Due to the frequent white light flares emitted by AU Mic, it was necessary to model both the flares and transits simultaneously in order to derive the most precise fits for the planet, AU Mic b. Furthermore, by allowing the time of the transits to vary in our fits, we determined that the transits do not show a constant linear ephemeris, and have a TTV amplitude of $\pm$ 80 seconds. We also saw two transits not associated with planet b. We fit these in conjunction with the tentative transit described by \citet{Plavchan2020} in Sector 1, and derived parameters for planet c, showing it to be a planet candidate with radius 2.79$ \pm 0.30$ R$_\oplus $. These results demonstrate the remarkable value of the 20-second light curves for gaining enhanced insight into both stellar magnetic activity and obtaining better understanding of planet parameters and dynamics.

\acknowledgments
This paper makes use of the 20-second cadence mode introduced in the TESS extended mission. We thank those on the TESS team who made this data collection mode possible.

This paper includes data collected by the TESS mission, which are publicly available from the Mikulski Archive for Space Telescopes (MAST). Funding for the TESS mission is provided by NASA's Science Mission directorate. We acknowledge the use of public TESS Alert data from pipelines at the TESS Science Office and at the TESS Science Processing Operations Center. 

This research has made use of the Exoplanet Follow-up Observation Program website, which is operated by the California Institute of Technology, under contract with the National Aeronautics and Space Administration under the Exoplanet Exploration Program. 

Resources supporting this work were provided by the NASA High-End Computing (HEC) Program through the NASA Advanced Supercomputing (NAS) Division at Ames Research Center for the production of the SPOC data products.

We are grateful to Jamie Lomax and John Wisniewski who kindly provided us with the HST spectrum used in this work.

E.A.G. thanks the LSSTC Data Science Fellowship Program, which is funded by LSSTC, NSF Cybertraining Grant \#1829740, the Brinson Foundation, and the Moore Foundation; her participation in the program has benefited this work. E.A.G. is thankful for support from GSFC Sellers Exoplanet Environments Collaboration (SEEC), which is funded by the NASA Planetary Science Division’s Internal Scientist Funding Model. The material is based upon work supported by NASA under award number 80GSFC21M0002. This work was also supported by NASA awards 80NSSC19K0104 and 80NSSC19K0315. 

This work was supported by grants to P.P. and L.V. from NASA (awards 80NSSC20K0251, 80NSSC21K0349, and 80NSSC21K0241), the National Science Foundation (Astronomy and Astrophysics grants 1716202 and 2006517), and the Mount Cuba Astronomical Foundation.

\facilities{TESS}

\software{
astropy \citep{exoplanet:astropy13,exoplanet:astropy18}, 
celerite2 \citep{celerite, exoplanet:foremanmackey18}, 
exoplanet \citep{exoplanet:exoplanet}, 
Forecaster \citep{forecaster}, 
IPython \citep{ipython}, 
Jupyter \citep{jupyer}, 
Lightkurve \citep{lightkurve}, 
Matplotlib \citep{matplotlib},
NumPy \citep{numpy}, 
PyMC3 \citep{exoplanet:pymc3}, 
STARRY \citep{exoplanet:luger18, exoplanet:agol20}, 
Theano \citep{exoplanet:theano}, 
xoflares \citep{xoflares}
TTV2Fast2Furious \citep{Hadden2018}, 
}

\clearpage

\bibliography{refs}{}
\bibliographystyle{aasjournal}

\end{document}